# Unsupervised MRI Reconstruction with Generative Adversarial Networks

Elizabeth K. Cole, John M. Pauly, *Member, IEEE,* Shreyas S. Vasanawala, Frank Ong

*Abstract*—Deep learning-based image reconstruction methods have achieved promising results across multiple MRI applications. However, most approaches require large-scale fully-sampled ground truth data for supervised training. Acquiring fully-sampled data is often either difficult or impossible, particularly for dynamic contrast enhancement (DCE), 3D cardiac cine, and 4D flow. We present a deep learning framework for MRI reconstruction without any fully-sampled data using generative adversarial networks. We test the proposed method in two scenarios: retrospectively undersampled fast spin echo knee exams and prospectively undersampled abdominal DCE. The method recovers more anatomical structure compared to conventional methods.

## I. INTRODUCTION

MAGNETIC resonance imaging (MRI) is an imaging modality which enables evaluation of soft-tissue anatomy and physiology. Unfortunately, MRI scans are intrinsically lengthy, which limits patient throughput and quality in uncooperative or young patients. The acquisition time can be reduced by undersampling in k-space. However, acquisition acceleration can result in nondiagnostic reconstructed images.

Many techniques exist clinically for improving the quality of these accelerated scans, which are based on parallel imaging (PI) (Griswold et al. 2002), (Pruessmann et al. 1999), and compressed sensing (CS) (Lustig et al. 2007). Recently, deep learning (DL) methods (Diamond et al. 2017 May 22; Chen et al. 2018; Hammernik et al. 2018; Yang et al. 2018; Eo et al. 2018; Cheng et al. 2018 May 8; Souza et al. 2019; Aggarwal et al. 2019; Mardani et al. 2019; Cole et al. 2020 Apr 3) have proven to be more powerful than traditional methods, providing more robustness, higher quality, and faster reconstruction speed. However, these techniques require a large number of fully-sampled acquisitions for supervised training. This poses a problem for applications such as dynamic contrast enhancement (DCE), cardiac cine, and 4D flow, where the collection of fully-sampled datasets is time-consuming, difficult, or impossible. As a result, DL-based methods often cannot be used in these applications.

There are two main possible ways to address this problem. First, parallel imaging-compressed sensing (PI-CS) reconstructions can be used as the ground truth for a DL framework (Cheng et al. 2018 May 8). However, the reconstructed images of the DL model are unlikely to be significantly better than PI-CS images. The second way is to formulate DL training by using only undersampled datasets for training (Lehtinen et al. 2018; Soltanayev and Chun 2018; Wu et al. 2018; Chen et al. 2019; Tamir et al. 2019; Zhussip et al. 2019; Yaman et al. 2019 Oct 20). However, such unsupervised training can be a difficult problem to solve and remains an active research topic.

A promising direction to address unsupervised MRI reconstruction is using generative adversarial networks (GANs) (Goodfellow et al. 2014b). GANs have proven very useful in creating visually appealing natural images (Zhu et al. 2016), modeling underlying data distributions (Goodfellow et al. 2014a; Radford et al. 2016), and constructing a generative model for supervised MRI reconstruction (Mardani et al. 2017 May 31; Yang et al. 2018; Mardani et al. 2019). Recently, Bora et al. (Bora et al. 2018) proposed a framework, called AmbientGAN, for learning generative models from underdetermined linear systems. They demonstrated encouraging results for small-scale simulated datasets, such as MNIST (LeCun Yann et al. 1998) and celebA (Liu et al. 2018). However, there are still several limitations when compared to supervised counterparts. In particular, AmbientGAN at inference time requires expensive iterative methods. Moreover, its generative models are stringent and cannot leverage unrolled network architectures, which are known to achieve state-of-the-art results on open reconstruction challenges (Knoll et al. 2020).

In this work, we propose a GAN training framework for learned MRI reconstruction that relies on only undersampled datasets and no fully-sampled datasets. The proposed method addresses the aforementioned limitations of AmbientGAN, enabling fast inference and high-quality reconstruction using unrolled networks. This enables efficient DL reconstruction when it is impossible or difficult to obtain fully-sampled data. In order to evaluate this method, we first implement the method on a retrospectively undersampled datasets and compare the

This paper was submitted for review on 08/29/20. This work was supported by NIH R01-EB009690, NIH R01-EB026136, and GE Healthcare.
Elizabeth K. Cole is with the Department of Electrical Engineering, Stanford University, CA 94301, USA (e-mail: ekcole@stanford.edu).
John M. Pauly is with the Department of Electrical Engineering, Stanford University, CA 94301, USA (email: pauly@stanford.edu).
Shreyas S. Vasanawala is with the Department of Radiology, Stanford University, CA 94301, USA (email: vasanawala@stanford.edu).
Frank Ong is with the Department of Electrical Engineering, Stanford University, CA 94301, USA (e-mail: franko@stanford.edu).

results to a supervised setting, but then implement the method on a prospectively undersampled set of DCE scans.

## II. RELATED WORK

Prior work on unsupervised learning has explored both unconditional and conditional GANs. Some work has only used traditional computer vision datasets, while other papers have applied algorithms to MRI data.

AmbientGAN (Bora et al. 2018) attempts to use only partial, noisy observations as training data to generate samples from a noise vector, using a GAN. The authors showed results in image inpainting, denoising and deconvolution. They were able to show that despite only training on such lossy data, their generator was still able to produce samples which recovered the underlying data distribution (Bora et al. 2018) of traditional computer vision datasets such as celebA and MNIST. This is similar to the problem we are trying to solve in unsupervised MRI reconstruction, where we only have undersampled k-space measurements for our training set. One notable difference, elaborated below, is that our reconstruction model is a conditional GAN where we directly learn a mapping from k-space to image domain. In contrast, AmbientGAN learns a network taking latent codes as inputs and outputs images.

Pajot et al. (Pajot et al. 2019), which is closely related to the work of (Bora et al. 2018), attempted to train an unsupervised model to recover an underlying signal from lossy observations. Again, the authors addressed a similar problem to the one we attempt to solve. However, the authors did not experiment on MRI data, opting again to only use traditional computer vision datasets.

Another work, called Noise2Noise (Lehtinen et al. 2018), has investigated training a model for noise removal based only on noisy training data. Here, the authors did experiment some with MRI data. However, one limitation of this work is that the authors did not accelerate the scans in a way that would be used clinically or in traditional MRI reconstruction. Additionally, the authors only trained on magnitude images, whereas state-of-the-art reconstruction techniques learn on the complex-valued data. The major limitation of that work was that it requires at least two independent scans of the same underlying image. Additionally, the authors did not incorporate the MR imaging physics-based model. A related paper by Huang et al. (Huang et al. 2020) used Noise2Noise and applied it to MRI, attempting to improve the work of Noise2Noise by enforcing data consistency. Various other works (Eldeniz et al. 2020; Liu, Eldeniz, et al. 2020; Liu, Sun, et al. 2020) also applied Noise2Noise to various MRI applications.

Another unsupervised method which uses data consistency for MRI reconstruction can be found in the work of Yaman et al. (Yaman et al. 2019 Oct 20). This method approximates actual data consistency using cross-validation. This work differs from the method described below in that it does not train a GAN, instead training only one network.

## III. METHODS

We consider the standard multi-channel Fourier acquisition model for MRI, which can be written as:

$$y = Ax + \varepsilon. \qquad (1)$$

where $y$ is the measured k-space data, $A$ is the imaging model, $\varepsilon$ is the additive complex Gaussian noise, and $x$ is the set of reconstructed images. The imaging model $A$ consists of data subsampling, a Fourier transform, and signal modulation by coil sensitivity maps $S$. We consider acquisitions with randomized sampling masks, such as the Poisson-disc variable density sampling, so $A$ is a random matrix drawn from a known distribution $p_A$. The overall k-space data distribution is denoted as $p_y$.

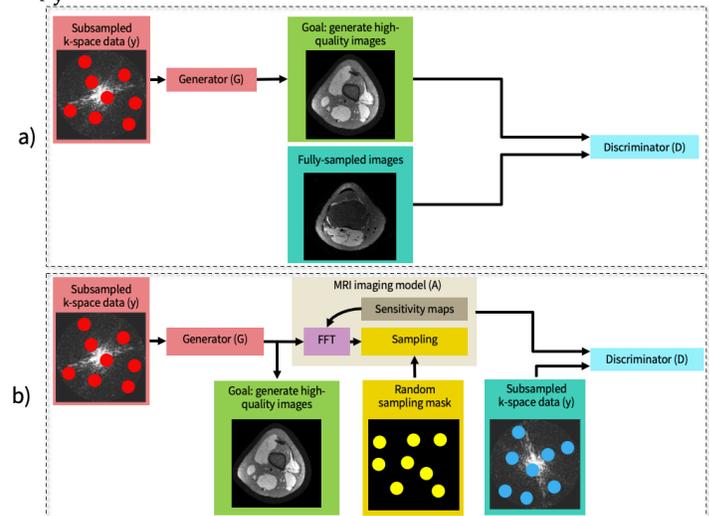

Fig. 1. A conventional supervised learning system (a) and our proposed unsupervised system (b).
(a) Framework overview in a supervised setting with a conditional GAN when fully-sampled datasets are available.
(b) Our proposed framework overview in an unsupervised setting. The input to the generator network is an undersampled complex-valued k-space data and the output is a reconstructed two-dimensional complex-valued image. Next, a sensing matrix comprised of coil sensitivity maps, an FFT and a randomized undersampling mask (drawn independently from the input k-space measurements) is applied to the generated image to simulate the imaging process. The discriminator takes simulated and observed measurements as inputs and tries to differentiate between them.

The generative adversarial network framework was first proposed by Goodfellow et al. (Goodfellow et al. 2014a). In this framework, a pair of neural networks are jointly trained. The generator network tries to map samples from a low-dimensional distribution that is easy to sample from (such as Gaussian noise) to samples from a high-dimensional space. Meanwhile, the discriminator network tries to differentiate between generated and real samples. To jointly train the networks, a min-max game is employed, where the loss function of the generator is based on the output of the discriminator. Convergence of a GAN is signified by obtaining equilibrium between the generator and discriminator. Many GAN loss functions exist, including that of the Wasserstein GAN (WGAN) (Arjovsky et al. 2017), the deep convolutional GAN (DCGAN) (Radford et al. 2016), and WGAN with gradient penalty (WGAN-GP) (Gulrajani et al. 2017 Mar 31). In this work, we use the losses of WGAN-GP because WGAN-GP has been shown to have potentially the best convergence (Gulrajani et al. 2017 Mar 31).

In a standard supervised setting, one could formulate a GAN for MRI reconstruction by using fully-sampled reconstructions as the real images (Shende et al. 2019). Figure 1a illustrates the overall framework of training a conditional GAN when fully-sampled datasets are available. However, when we only have access to training datasets where $A$ is underdetermined due to data subsampling, we don't have a clean ground truth to use as those real images.

AmbientGAN attempts to solve this problem of lack of clean training data by training the discriminator to differentiate between a real measurement from a simulated measurement of a generated image. In the context of MRI reconstruction, the AmbientGAN objective is:

$$\min_G \max_D \mathbb{E}_{y \sim p_y}[q(D(y))] + \mathbb{E}_{z \sim p_z, A \sim p_A}[q(1 - D(A(G(z))))]. \quad (2)$$

where $G$ is the generator network, and $D$ is the discriminator network. $y$ denotes the observed subsampled measurements drawn from the distribution $p_y$. The vector $z$ denotes a random latent vector of a distribution $p_z$ that is easy to sample from, such as IID Gaussian noise. $q(\cdot)$ denotes the quality function used to define the objective. For vanilla GAN, $q(t) = log(t)$, and for the WGAN (Arjovsky et al. 2017) and WGAN-GP (Gulrajani et al. 2017 Mar 31), $q(t) = t$. The generated images are given by $G(z)$ and generated measurements are given by $A(G(z))$.

Note that in AmbientGAN, the input vector $z$ is mapped from a distribution of random latent vectors to a higher dimensional image space. However for reconstruction, we want a generative mapping from k-space to an image space, not from a latent distribution to an image space. Therefore, we propose to use a conditional GAN, where the input is actually subsampled k-space data $y$ from the distribution $p_y$, not the noise vector $z$.

The resulting reconstruction of our model is non-iterative, which is a major advantage of our method. $G(y)$ is the resulting generated high-resolution image from the input undersampled k-space data. In contrast, the AmbientGAN setup is unable to do a non-iterative reconstruction because the latent code must be solved during each inference step.

Concretely, our overall proposed objective is:

$$\min_G \max_D \mathbb{E}_{y \sim p_y}[q(D(y))] + \mathbb{E}_{y \sim p_y, A \sim p_A}[q(1 - D(A(G(y))))]. \quad (3)$$

where the discriminator is trained to distinguish between real measurements $y$ and generated measurements $A(G(y))$.

A subtle difference between the AmbientGAN setting and our setting is that all of our variables are in the complex domain instead of real. However, this can be dealt with by splitting the real and imaginary components into two channels, or using custom complex-valued building blocks (Cole et al. 2020 Apr 3).

One potential concern with this objective is that the generator may directly output the zero-filled reconstruction of $y$ because both generator and discriminator take undersampled k-space measurements as inputs. The key to addressing this problem is that the simulated imaging model $A$ in the objective is independently drawn with respect to the generator inputs $y$. Therefore, the simulated measurements $A(G(y))$ are unlikely to be sampled at the same k-space locations as the input $y$. The discriminator would likely enforce the generator to fill in missing k-space measurements, because otherwise the discriminator can easily classify the generated data as fake.

The proposed unsupervised framework is shown in Figure 1b. The input to the generator network is an undersampled complex-valued k-space data and the output is a reconstructed two-dimensional complex-valued image. Next, a sensing matrix comprised of coil sensitivity maps, an FFT and a randomized undersampling mask (drawn independently from the input k-space measurements) is applied to the generated image to simulate the imaging process. The discriminator takes simulated and observed measurements as inputs and tries to differentiate between them.

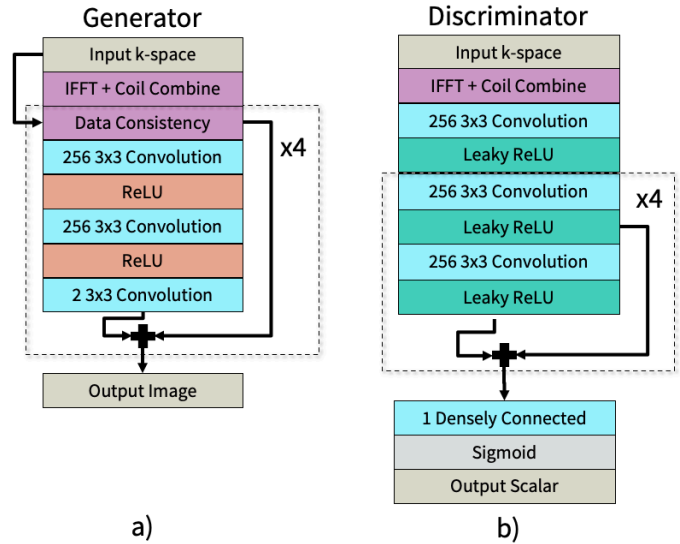

Fig. 2a. The generator architecture, which is an unrolled network based on the Iterative Shrinkage-Thresholding Algorithm and includes data consistency.

Fig. 2b: The discriminator architecture, which uses leaky ReLU in order to backpropagate small negative gradients into the generator.

### A. Network Details

The WGAN-GP (Gulrajani et al. 2017) quality and penalty functions are used in the training objective. An unrolled network (Diamond et al. 2017 May 22) based on the Iterative Shrinkage-Thresholding Algorithm (ISTA) (Beck and Teboulle 2009) is used as the generator architecture, shown in Figure 2a. The unrolled network is a common data-driven approach to reconstruction which also incorporates known MR physics (Sandino et al. 2018; Cheng et al. 2018 May 8; Mardani et al. 2018 Jun 1; Tamir et al. 2019; Lei et al. 2019 Oct 15; Yaman et al. 2019 Oct 20; Cole et al. 2020; Liang et al. 2020; Sandino et al. 2020). The discriminator architecture, based on a simple convolutional neural network with residual structure, is shown in Figure 2b. Residual connections help to enforce the original structure of the input image. Leaky ReLU is used as the final activation function instead of the traditional ReLU in order for the discriminator to be able to backpropagate small negative gradients into the generator.

### B. Dataset Details

Two datasets were obtained with Institutional Review Board (IRB) approval and subject informed consent. The first dataset was a set of fully sampled 3T knee images acquired using 8 channel coil arrays and a 3D FSE CUBE sequence with proton density weighting including fat saturation (K. Epperson, A. M. Sawyer, M. Lustig, M. T. Alley, M. Uecker, P. Virtue, P. Lai and Vasanawala 2013). 15 subjects were used for training and 3 subjects were used for testing. The readout was in the superior/inferior direction, making that direction a natural choice to remain fully-sampled. Therefore, we subsample in the left/right and anterior/posterior directions. Each subject had a complex-valued volume of size 320 x 320 x 256 that was split into axial slices of size 320 x 256. Because a fully-sampled ground truth exists for this scenario, we can quantitatively validate our results. We created undersampled images by applying pseudo-random Poisson-disc variable-density sampling masks to the fully-sampled k-space. Although we initially use fully-sampled datasets to create sub-sampled datasets, it is critical to note that the generator and discriminator are never trained with fully-sampled data.

The second dataset consists of dynamic contrast enhanced (DCE) acquisitions of the abdomen, with a fat-suppressed butterfly-navigated free-breathing SPGR acquisition (Zhang et al. 2015). 886 subjects were used for training and 50 subjects were used for testing. It is impossible to obtain fully-sampled data for DCE because the dynamics of the intravenously injected contrast are faster than can be captured at full sampling by the imaging hardware. Each scan acquired a volumetric image with dimensions of 192 x 180 x 80. The raw data was compressed from 32 channels to 6 virtual channels using a singular-value-decomposition-based compression algorithm (Zhang et al. 2013). Data were fully sampled in the $k_x$ direction (spatial frequency in x) and were subsampled in the $k_y$ and $k_z$ directions. Images were subsampled with a total acceleration factor of around 5.

The Berkeley Advanced Reconstruction Toolbox (BART) (Tamir et al. 2016) was used to estimate sensitivity maps, generate undersampling masks, and perform a compressed sensing reconstruction of these datasets for comparison purposes. Coil sensitivity maps was generated using ESPIRiT (Uecker et al. 2014).

## IV. EXPERIMENTS

### A. Retrospectively Undersampled Knee Dataset

First, we trained two GANs, one supervised, and one using our unsupervised method, on the set of knee scans. We did this to quantitatively evaluate the reconstruction performance gap between a traditional supervised GAN and our proposed unsupervised method. We also compared our proposed unsupervised method to compressed sensing with $L_1$-wavelet regularization, another reconstruction method which requires no fully-sampled data, and which is routinely used in our clinical practice. For each knee scan, we used a fully-sampled calibration region of 20 × 20 in the center of k-space. To compute image quality, we evaluated average normalized root-mean-square error (NRMSE), peak signal-to-noise ratio (PSNR), and structural similarity index (SSIM) (Wang et al. 2004) between the reconstructed image and the fully-sampled ground truth on test datasets. The generator and discriminator of both of these GANs were trained with 256 feature maps. Both generators had 4 residual blocks and 5 iterations.

Next, we evaluated the reconstruction performance on the set of knee scans of the unsupervised GAN as a function of the acceleration factor of the training datasets. We compared this to the reconstruction performance of the supervised GAN, trained on fully-sampled datasets. For this experiment, we used a calibration region of 5 x 5 in the center of k-space.

### B. Prospectively Undersampled DCE Dataset

Finally, we trained our unsupervised GAN on the set of abdominal DCE scans. Because DCE must be undersampled for adequate temporal resolution, we have no ground truth to quantitatively assess reconstruction performance. Instead, we compare to a CS (Lustig et al. 2007) reconstruction that is used in our routine clinical practice, and qualitatively evaluate the sharpness of the vessels and other anatomical structures in the generated images. Both the network and CS were done frame-by-frame. The generator and discriminator of this GANs were each trained with 512 feature maps. The generator had 4 residual blocks and 5 iterations of the unrolled network. We compared the inference time per slice between CS and our unsupervised GAN.

The number of feature maps, unrolled iterations, and residual blocks were chosen for each model to maximize the computational capacity of the network, and thus maximize the reconstruction quality. Many authors have shown that as the size of a network increases, so does its performance (Nakkiran et al. 2019 Dec 4). Cheng et al. specifically showed that for unrolled network architectures, as the number of iterations increased, the reconstructions' PSNR and SSIM increased, while the NRMSE decreased (Cheng et al. 2018 May 8). Therefore, we chose the biggest model that would fit on our GPUs.

All networks were trained with a batch size of one and optimized with the Adam optimizer (Kingma and Ba 2015) with $\beta_1 = 0.9$, $\beta_2 = 0.999$, and a learning rate of 1e-8. In all networks, the generator and discriminator were trained for one iteration per training step. Networks were trained using an NVIDIA Titan Xp graphics card and an NVIDIA GeForce GTX 1080 Ti graphics card. The proposed methods were implemented in Python using Tensorflow.

In the spirit of reproducible research, we provide a software package in Tensorflow to reproduce the results described in this article: https://github.com/ekcole/unsupGAN-release.

## V. Results

### A. Retrospectively Undersampled Knee Dataset

Figure 3 displays a comparison of PSNR, SSIM, and NRMSE between reconstructions from CS with $L_1$-wavelet regularization, our proposed unsupervised GAN, and a standard supervised GAN on our test dataset. The proposed unsupervised GAN had superior PSNR, NRMSE, and SSIM compared to the CS reconstruction. Additionally, the proposed unsupervised GAN only had 0.78% worse PSNR, 4.17% worse NRMSE, and equal SSIM compared to the supervised GAN. The error bars on each series represent the standard deviation of each image metric. The error bars of CS are much larger than the error bars of the unsupervised and supervised GANs. This suggests that the unsupervised generative method is more stable than CS.

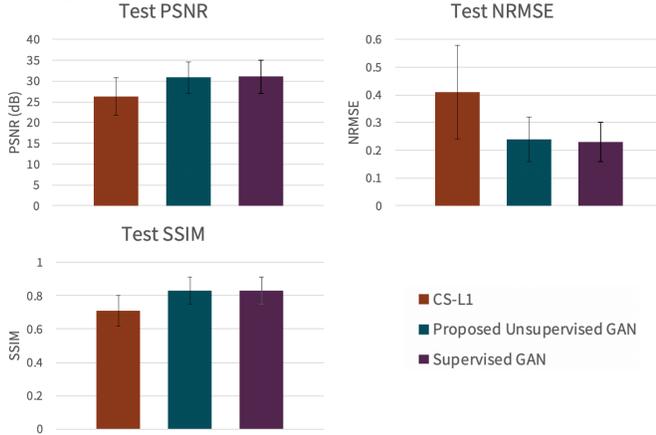

Fig. 3. Image metrics calculated on test datasets with reconstructions from CS with $L_1$-wavelet regularization, the proposed unsupervised GAN, and the supervised GAN. The error bars on each series represent the standard deviation of each image metric. The error bars of CS are much larger than the error bars of the unsupervised and supervised GANs. This shows that our unsupervised generative method may more stable than CS.

Representative results in the knee scenario are shown in Figure 4. The columns represent, from left to right, the input undersampled complex image to the generator, the output of the unsupervised generator, the output of the supervised generator, and the fully-sampled image. The acceleration factors of the input images are 15.6, 6.5, and 9.9, from top to bottom. In all rows, the unsupervised GAN has superior PSNR, NRMSE, and SSIM compared to CS. In the first row, the unsupervised GAN has metrics that are worse than the supervised GAN. In the middle and last rows the unsupervised GAN has metrics that come relatively close to the performance of the supervised GAN. In the unsupervised GAN, the generator markedly improves the image quality by recovering vessels and structures that were not visible before, but uses no ground truth data in the training.

The results of the reconstruction performance on the set of knee scans of the unsupervised GAN as a function of the acceleration factor of the training datasets is shown in Figure 5. The supervised GAN uses fully-sampled training datasets, and is plotted for reference. CS does not use training datasets, and is also plotted for reference. As the acceleration factor of the training data is increased, the reconstruction performance, measured by PSNR, NRMSE, and SSIM, decreases. The biggest drop-off in performance occurs between training acceleration factors of 8 and 10. However, the reconstruction performance is still competitive with that of the supervised GAN. Interestingly, SSIM is not monotonic in acceleration.

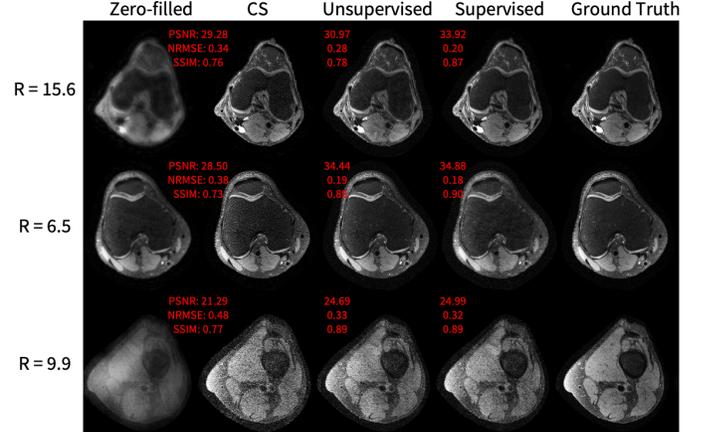

Fig. 4. Knee application representative results, showing, from left to right: the input undersampled complex image to the generator, the output of the unsupervised generator, the output of the supervised generator, and the fully-sampled image. The acceleration factors of the input image are 15.6, 6.5, and 9.9, from top to bottom. In all rows, the unsupervised GAN has superior PSNR, NRMSE, and SSIM compared to CS. In the first row, the unsupervised GAN has metrics that are notably worse than the supervised GAN. In the middle row and last rows, the unsupervised GAN has metrics that come close to the performance of the supervised GAN.

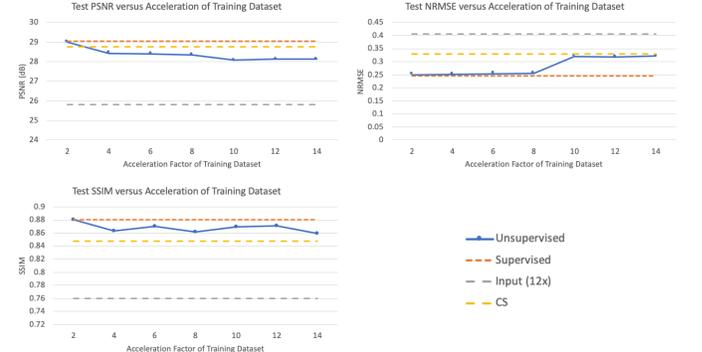

Fig. 5. The results of the reconstruction performance on the set of knee scans of the unsupervised GAN as a function of the acceleration factor of the training datasets. The y-axis represents PSNR, NRMSE, or SSIM, depending on the plot. The x-axis represents the acceleration factor of the unsupervised GAN's training datasets. The supervised GAN uses fully-sampled training datasets, and CS does not use training datasets. Therefore, neither of their performance varies with the x-axis. They are both plotted for reference. As the acceleration factor of the training data is increased, the reconstruction performance of the unsupervised GAN, measured by PSNR, NRMSE, and SSIM, decreases. The biggest drop-off in performance occurs between training acceleration factors of 8 and 10. However, the reconstruction performance is still competitive with that of the supervised GAN.

### B. Prospectively Undersampled DCE Dataset

Representative DCE results are shown in Figure 6. The leftmost column is the input zero-filled reconstruction, the middle column is our generator's reconstruction, and the rightmost column is the CS reconstruction. The generator greatly improves the input image quality by recovering sharpness and adding more structure to the input images. Additionally, the proposed method produces a sharper reconstruction compared to CS. In the first row, the anatomical right kidney (left side of image) of the unsupervised GAN is visibly much sharper than that of the input and CS. A video of all 18 phases of these images is attached in the Supplementary Information.

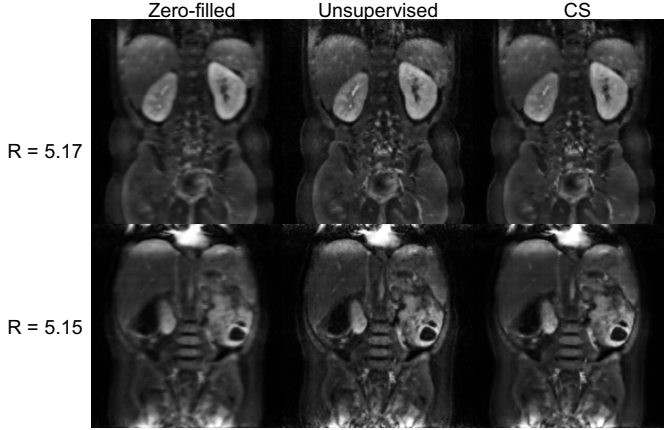

Fig. 6. 2D DCE application representative results, where the left slice is the magnitude of one input undersampled complex image to the generator, the middle slice is the output of the generator and the right slice is a compressed sensing L1-wavelet regularization. The generator greatly improves the input image quality by recovering sharpness and adding more structure to the input images. Additionally, the proposed method produces a sharper reconstruction compared to CS. In the first row, the kidneys of the unsupervised GAN is visibly much sharper than that of the input and CS. A video of all 18 phases of these images is attached in the Supplementary Information.

A comparison of the average inference time per two-dimensional DCE slice between CS and our unsupervised GAN is shown in Figure 7. The error bars of the figure represent the calculated standard deviation of the inference time. The inference time of our unsupervised method is approximately 7 times faster than CS. The standard deviation of the inference time for each method are approximately equivalent.

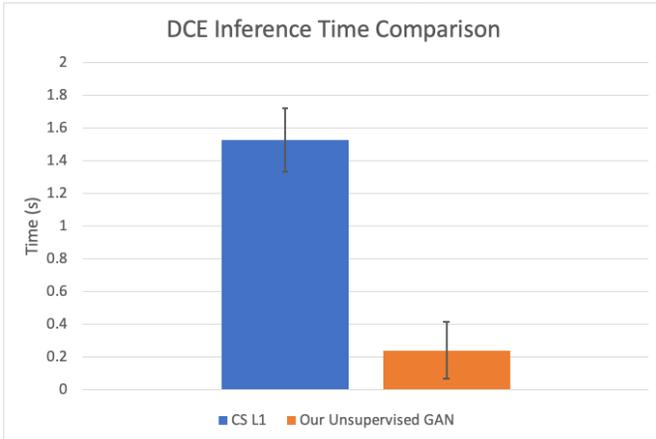

Fig. 7. Mean reconstruction speed comparison per each two-dimensional DCE test slice. The inference time of our unsupervised method is approximately 7 times faster than CS. Error bars represent the standard deviation. The standard deviation of the inference time for each method are approximately equivalent.

## VI. DISCUSSION

### A. Recap

In this work, we have developed a generative model for unsupervised MRI reconstruction. This method differs the current thrust of MRI reconstruction work by obviating any fully-sampled data as ground truth for training. Our method is different from other unsupervised MRI reconstruction work because we choose to train a generative adversarial model for reconstruction, instead of only one network.

In the knee dataset, the reconstructions from our proposed unsupervised method achieve superior SSIM, PSNR, and NRMSE compared to a CS reconstruction. Additionally, the gap between the proposed unsupervised GAN and the supervised GAN is fairly small, with a difference of 0.78% worse PSNR, 4.17% worse NRMSE, and equal SSIM.

As the acceleration factor of the training knee datasets is increased, the reconstruction performance, measured by PSNR, NRMSE, and SSIM, decreases. This trend is as one would expect because as the acceleration factor of the training dataset increases, the GAN has less range of sampled k-space to learn from.

In the DCE application, the generated images are sharper than those reconstructed by. Additionally, the inference time of our method is much faster than CS.

Through our results, we have demonstrated that when fully-sampled data is available, supervised training should still be used for best reconstruction quality. However, in the situations where fully-sampled data is not available for training a reconstruction model, our unsupervised method can still produce reconstructions which are comparable to a supervised counterpart and better than CS. Although we have demonstrated an application in DCE, which is common across a range of oncologic imaging indications, these concepts can potentially be leveraged for higher dimensional acquisitions in cardiac imaging, such as volumetric cine and 4D flow, as well as in neurologic imaging, such as DTI and fMRI.

### B. Advantages

The main advantage of this method over existing DL reconstruction methods is the obviation of fully-sampled data. Another benefit is that other additional dataset is needed to use as ground-truth, as in some other works on semi-supervised training (Lei et al. 2019 Oct 15). Additionally, the method produces better quality reconstruction compared to baseline CS methods.

This method is generalizable, and could be easily extended to other GAN losses, such as WGAN (Arjovsky et al. 2017) or DCGAN (Radford et al. 2016), and network architectures, such as variational networks (Hammernik et al. 2018), U-Nets (Ronneberger et al. 2015), or hybrid-domain networks (Eo et al. 2018). This technique can be applied to many different dimensionalities and applications, and can thus be demonstrated for 2D slices, 3D volumes, 4D datasets, and 2D slices plus a time dimension.

Additionally, this method could also be useful for high noise environments where the acquisition of high SNR data is

7difficult. Other adverse situations where ground truth data are precluded include real-time imaging due to motion and arterial spin labeling due to low SNR. Further applications where it is hard to fully sample are time-resolved MR angiography, cardiac cine, low contrast agent imaging, EPI-based sequences, diffusion tensor imaging, and fMRI.

Outside of MRI, this method can potentially have applications in areas where obtaining fully-sampled data is difficult or impossible, such as dynamic PET (Gong et al. 2019) or computed tomography (CT) (Gallegos et al. 2018).

### C. Limitations

One limitation of this method is that because our framework uses a GAN, we have to train two separate neural networks, which can take more memory and time. Additionally, another challenge is that the training of the generator and discriminator must be balanced, so that they don't become unstable. This can potentially be done by tuning the number of iterations the discriminator and generator are trained per training step to balance both networks. Also, the model sizes and parameters could be optimized for each dataset, although this was not the focus of this work.

Another challenge of this method is that the imaging model must be correctly specified to simulate k-space measurements. This is straightforward for the considered applications with Cartesian sampling, but is far more difficult in applications with system imperfections and motion corruption.

Because no fully-sampled data existed for our DCE dataset, it was difficult to quantitatively validate the DCE experiments. In the future, using a digital or real phantom with dynamic contrast could provide a more accurate assessment of various unsupervised reconstruction methods. This could improve the experiments. However, phantoms often cannot capture the complexity of real people.

In the future, image quality could potentially be improved by adding some kind of perceptual loss to the loss function of the generator, such as a total variation loss (Rudin et al. 1992) of the generated image, a feature reconstruction loss between the generated and real images (Johnson et al. 2016), or a style reconstruction loss between the generated and real images (Johnson et al. 2016).

## VII. Conclusion

In this work, we propose an unsupervised GAN framework for reconstruction without a ground truth. We show that the proposed method outperforms existing traditional methods such as compressed sensing. Our proposed method has NRMSE, PSNR, and SSIM values which come close to the performance of a supervised GAN. In contrast to most deep learning reconstruction techniques, which are supervised, this method does not need any fully-sampled data. With the proposed method, accelerated imaging and accurate reconstruction can be performed in applications in cases where fully-sampled datasets are difficult to obtain or unavailable.